\documentclass[%
reprint,
superscriptaddress,
amsmath,amssymb,
aps,
prstab,
]{revtex4-2}
\usepackage{xcolor}
\usepackage{graphicx}
\usepackage{dcolumn}
\usepackage{bm}
\usepackage{booktabs}
\usepackage{multirow}
\usepackage{amssymb}
\usepackage{float}
\usepackage{epstopdf}

\begin{document}

\title{Oscillatory magnetic field-dependent critical temperatures of ultraclean Type-II superconductors}
\author{Aiying Zhao}
\email{ayzhao0909@sina.cn}
\affiliation{Institute of Theoretical Physics, University of Science and Technology Beijing, Beijing 100083, People’s Republic of China}
\author{Richard A Klemm}
\email{richard.klemm@ucf.edu}
\affiliation{Department of Physics, University of Central Florida, Orlando, FL 32816-2385, United States of America}
\author{Qiang Gu}
\email{qgu@ustb.edu.cn}
\affiliation{Institute of Theoretical Physics, University of Science and Technology Beijing,  Beijing 100083, People’s Republic of China}

\date{\today}

\baselineskip12pt

\begin{abstract}
The influence of the Zeeman energy and the Landau levels (LLs) arising from an applied  magnetic field ${\bf B}$ upon the critical temperature $T_c$ is studied using a fully quantum mechanical method within the framework of the Bardeen–Cooper–Schrieffer (BCS) theory of superconductivity that forms from an ultraclean metal. As in semiclassical treatments, we  found that two electrons can form Cooper pairs  with opposite spins and momenta in the  ${\bf B}$ direction while either in the same or in neighboring LLs. However, the fully quantum mechanical treatment of the LLs causes $T_c({\bf B})$ for electrons  paired on the same LL to oscillate about the critical temperature of the BCS theory, similar to that of the de Haas-van Alphen effect. The Zeeman energy causes $T_c({\bf B})$ to decrease in an oscillatory fashion with  increasing   ${\bf B}$ for 
electrons paired either on the same or on neighboring LLs. For the Zeeman $g > 1$, pairing on neighboring LLs results in the highest $T_c({\bf B})$. For $g < 1$, pairing on the same LLs gives the highest $T_c({\bf B})$. In addition,  $T_c({\bf B})$ for electrons paired on  neighboring LLs exhibits an apparent symmetry around  $g=2$, as 
the oscillatory critical temperature behaviors are nearly identical for $g=2\pm\delta$. 
\end{abstract}

\pacs{05.20.-y, 75.10.Hk, 75.75.+a, 05.45.-a} \vskip0pt

\maketitle

\section{Introduction}

The critical temperature is a fundamental and paramount characteristic of superconductors. In the quest to understand superconductivity physics, both experimental measurements 
and theoretical investigations of the critical temperature $T_c$ are crucial components of it. Recent advancements in experimental technology have significantly contributed to 
progress in superconductivity research \cite{Hemely-2019,Eremets-2019,Eremets-2015,Mozaffari-2019,Minkov-2022}. The quality and purity of the materials have been consistently improving. Various novel phenomena of superconductors have been observed in the presence of magnetic fields ${\bf B}$, such as the upper critical fields parallel to the conducting  planes of two-dimensional materials, which often violate the Pauli limit \cite{Ebling-2014,MoS2-2016,Falson-2020,Cao-2018}, and the reentrance of superconductivity in magnetic fields higher than the nominal upper critical field \cite{Cao-2021,Ran-2019,Ran-2023}. In addition, bi- or trilayer combinations of ferromagnetic and superconducting layers and magnetically coupled mesoscopic loops have exhibited oscillations in $T_c({\bf B})$ \cite{Bergeret-2005,Morelle,Buzdin-2005}. Investigating the quantum correlations between the superconducting critical temperature and the external magnetic field is therefore very important.
  
In theoretical research, the semiclassical phase approximation has been a well-established and widely used method for exploring the effects of magnetic fields on the upper critical field and the critical temperature of superconductors \cite{WH-1966,GG-1966,GG-1968,Rajagopal-1966,KBL,Tesanovic-1989}. This approximation assumes that the influence of the applied magnetic field on a moving electron is solely represented by a phase factor, which multiplies the wave function of the free electron by 
$\exp({\frac{ie}{\hbar}\int_{\bm{r}_1}^{\bm{r}_2}\bm{A}\cdot d\bm{r}})$. This assumption is valid in weak magnetic fields, and the magnitude of the wave function of Cooper pairs, or the gap, is given by $\Delta(\rho,z)=\exp(-\frac{eB}{2}\rho^2)$, in which $\rho=(x,y)$ and ${\bm r}=(\rho,z)$. This linearized solution indicates that the Cooper pairs occupy the lowest LLs. However, in the presence of a strong magnetic field and near to the superconducting-to-normal state transition, it is challenging to maintain the assumption that electrons forming Cooper pairs behave as free electrons, rather than being quantized into LLs. If electrons forming Cooper pairs are indeed quantized, the density of states and the number of LLs near the Fermi surface (FS) become dependent upon the magnetic field. Consequently, the critical temperature of ultraclean superconductors in the presence of a magnetic field should exhibit oscillatory behavior similar to that observed in the de Haas-van Alphen effect.  

The Zeeman energy, which represents the response of an electron spin to a magnetic field, can play a crucial role in influencing the upper critical field of Type-II 
superconductors. One of the most significant consequences of the Zeeman energy is the determination of the Pauli limit, $B_p$ \cite{Chandrasekhar-1962,Clogston-1962}.  The Pauli limit is used to distinguish between different pairing configurations of the spin components in superconducting materials and to predict the irreversible first-order phase transition in spin-singlet superconductors, where the Zeeman splitting of the electronic spin state energy exceeds  the energy of the superconducting gap \cite{Sarma-1963}.  Therefore, one must carefully treat the influence of the Zeeman energy upon the critical temperature of a superconductor.

Here we explore the influences of the Zeeman energy and the LLs on the critical temperature of an ultraclean superconductor in the presence of a magnetic field using a fully quantum mechanical approach. We express the energy of the electrons in an external magnetic field as the quantized energy levels in the direction perpendicular to the magnetic field. When the Zeeman energy is included, the quantized energy for electron pairs with opposite spins in the same LL is split into 2 values. This splitting gives the possibility of electrons forming Cooper pairs on the same or on neighboring (and potentially on next-neighboring, etc.) LLs with opposite spins and momenta in the direction of the magnetic field. We present our $T_c(B)$ behavior on a much finer scale than for macroscopic ferromagnetic and superconducting layers, requiring ultraclean samples to observe the oscillations experimentally.  One possible material could be ultrapure Al films, which have been shown to be Type-II  superconductors for rather thick films \cite{Nsanzineza,Lopez-Nunez}.  We argue that the Cooper pairs described in our present study differ from those in the earlier novel states proposed for high-temperature superconductors, low-carrier-density semimetals, and semiconductors in high magnetic fields \cite{Tesanovic-1989,Tesanovic-1991,Rasolt}.  These earlier states lack the Meissner effect and exhibit quasi-one-dimensional behavior.  In contrast, our work suggests that ultrapure conventional superconductivity can persist in high magnetic fields, even surpassing the Pauli limit.  We investigated microscopically  the influence of various pairing possibilities due to the applied magnetic field upon the critical temperature. 

This paper is organized as follows. The second section introduces the quantized Hamiltonian for electrons paired on the same and on neighboring LLs, and derives the critical 
temperatures for both scenarios. The third section presents and discusses numerical results based on the equations from the second section, along with details of the effects of the LLs and the Zeeman energy. Finally, the main findings of this study are summarized in the fourth section.

\section{Model}
\subsection{Electrons paired on the same Landau level}

\begin{figure}
\centering
{\includegraphics[width=0.45\textwidth]{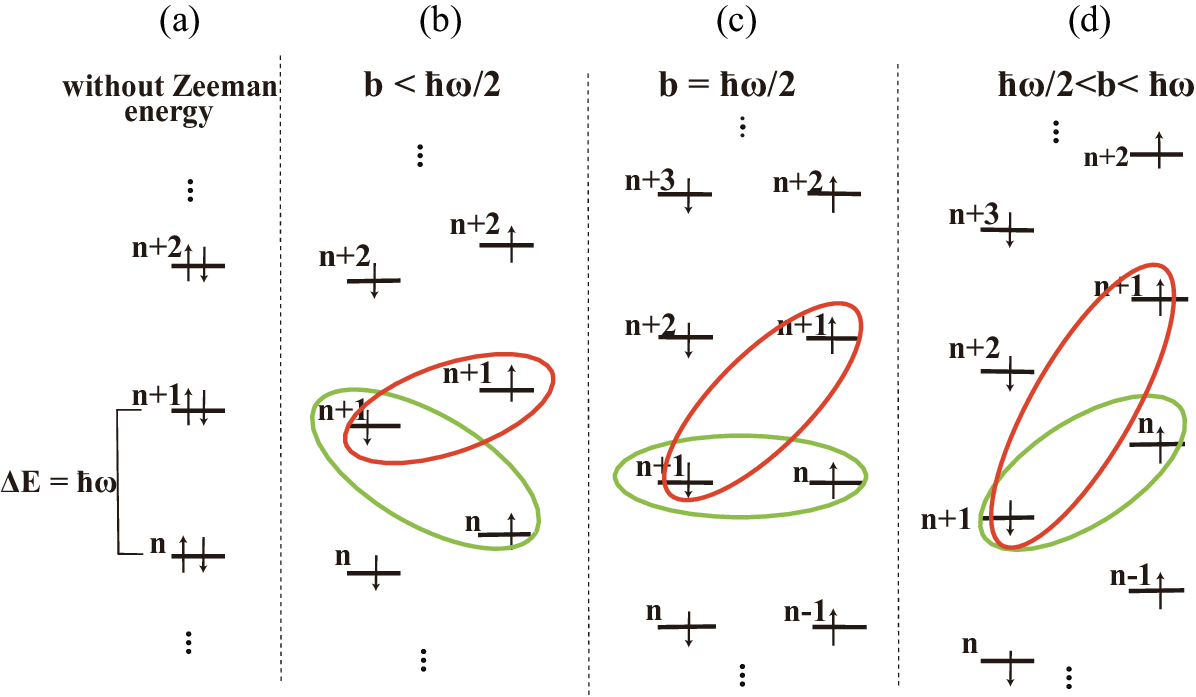}
\caption{ A schematic diagram of LLs and electron pairing. The arrows depict the electron spins, $n$, $n+1$... represent the number of LLs in a thin shell around the FS, 
$b$ is the Zeeman energy, $\hbar\omega$ is the energy difference between the nearest neighboring LLs. The first column (a) doesn't include the Zeeman energy, and the last three columns (b-d) show three different cases of the Zeeman energy relative to one half of the energy difference between the neighboring LLs. The red circles in (b-d) illustrate electrons paired on the same LLs, whereas the green circles represent electrons paired on  neighboring LLs. In the figures, the energy of electrons with up spin in the $n$th LL is $n\hbar\omega+b$, and the energy of electrons with down spin electrons in the $(n+1)$th LL is $(n+1)\hbar\omega-b$. Figure (c) specifically depicts the case where electrons paired on the neighboring LLs have equal energy.}
\label{fig1}}
\end{figure}

The Hamiltonian for electrons on an ellipsoidal Fermi surface paired on the same LL is
\begin{eqnarray}
H&=&\sum_{n=0}^{n_{max}} \frac{eB L^{2} L_z}{2\pi \hbar} \int \frac{d k_{z}}{2\pi}
 \biggl( (\epsilon_{n,k_{z}}+b ) a^{\dagger}_{n,k_{z}\uparrow}a_{n,k_{z}\uparrow} \nonumber\\
 &&+(\epsilon_{n,k_{z}}-b ) a^{\dagger}_{n,-k_{z}\downarrow}a_{n,-k_{z}\downarrow}
 \biggr)
 -V_{int}\sum_{n=0}^{n_{max}}  \frac{eB L^{2}L_z}{2\pi \hbar} \nonumber \\ 
 &&\int \frac{d k_{z}}{2\pi}cite
    (a^{\dagger}_{n,k_{z}\uparrow}a^{\dagger}_{n,-k_{z}\downarrow}  a_{n,-k_{z}\downarrow}a_{n,k_{z}\uparrow}),
\end{eqnarray}
in which $-V_{int}$ is a small attractive electron-electron interaction for electronic energies within a shell around the FS, $\epsilon_{n,k_z} = 
E_z+(n+\frac{1}{2})\hbar\omega-\mu_F$, $E_z=\frac{\hbar^2 k_{z}^{2}}{2m_{z}}$, $\hbar$=$\frac{h}{2\pi}$ is the reduced Planck constant,  $\omega=\frac{eB}{m_{xy}}$ is the cyclotron frequency, $m_{z}$ and $m_{xy}=\sqrt{m_x^2+m_y^2}$ represent the effective masses parallel and perpendicular to the magnetic field on the ellipsoidal Fermi surface \cite{KC},  $b=|\frac{g}{2} \mu_B \bm{\sigma}\cdot \bm{B}| $ is the Zeeman energy, $\frac{eBL^2}{2\pi\hbar}$ is the degeneracy of a  LL,  $L^2$ is the area of the sample normal to ${\bf B}$, and $L_z$ is the thickness of the sample along ${\bf B}$.

To diagonalize the Hamiltonian, we employ a modified Bogoliubov-Valatin transformation by defining the new fermionic operators as
\begin{align}
  \gamma_{n,k_z,\uparrow}&=\mu_{k_z}a_{n,k_z,\uparrow}-\nu_{k_z}a^{\dagger}_{n,-k_z,\downarrow},\nonumber\\
  \gamma_{n,-k_z,\downarrow}&=\mu_{k_z}a_{n,-k_z,\downarrow}+\nu_{k_z}a^{\dagger}_{n,k_z,\uparrow}; \nonumber\\
  \gamma^{\dagger}_{n,k_z,\uparrow}&=\mu^{\ast}_{k_z}a^{\dagger}_{n,k_z,\uparrow}-\nu^{\ast}_{k_z}a_{n,-k_z,\downarrow},\nonumber\\
  \gamma^{\dagger}_{n,-k_z,\downarrow}&=\mu^{\ast}_{k_z}a^{\dagger}_{n,-k_z,\downarrow}+\nu^{\ast}_{k_z}a_{n,k_z,\uparrow},
\end{align}
where the coefficients $\mu_{k_z}$, $\nu_{k_z}$ satisfy
\begin{align}
  |\mu_{k_z}|^2+|\nu_{k_z}|^2=1.
\end{align}

Then it is easy to obtain
\begin{align}
  a_{n,k_{z},\uparrow}&=\mu^{\ast}_{k_z} \gamma_{n,k_z,\uparrow} + \nu_{k_z}\gamma^{\dagger}_{n,-k_{z},\downarrow},\nonumber\\
  a_{n,-k_{z},\downarrow}&=\mu^{\ast}_{k_z}\gamma_{n,-k_z,\downarrow}-\nu_{k_z}\gamma^{\dagger}_{n,k_z,\uparrow}; \nonumber\\
  a^{\dagger}_{n,k_{z},\uparrow}&=\mu_{k_z} \gamma^{\dagger}_{n,k_z,\uparrow} + \nu^{\ast}_{k_z}\gamma_{n,-k_{z},\downarrow},\nonumber\\
  a^{\dagger}_{n,-k_{z},\downarrow}&=\mu_{k_z}\gamma^{\dagger}_{n,-k_z,\downarrow}-\nu^{\ast}_{k_z}\gamma_{n,k_z,\uparrow}.
\end{align}

After substituting equation (4) into equation (1), the Hamiltonian becomes
\begin{align}
 H=H_0+H_1+E_g,
\end{align}
where
\begin{align}
  H_0&= \frac{eBL^2 L_z}{2\pi \hbar}  \sum_{n=0}^{n_{max}} \int \frac{d k_z}{2\pi}
 \Biggl[ \biggl(  (\epsilon_{n,k_z}+b )|\mu_{k_z}|^2 \nonumber\\
  &-(\epsilon_{n,k_z}-b)|\nu_{k_z}|^2+ \mu^{\ast}_{k_z} \nu_{k_{z}}\Delta^{\ast}\nonumber\\
  & + \mu_{k_z} \nu^{\ast}_{k_z} \Delta    \biggr)
  \gamma^{\dagger}_{n,k_z,\uparrow} \gamma_{n,k_z,\uparrow}  \nonumber\\
  & +\biggl(  -(\epsilon_{n,k_z}+b )|\nu_{k_z}|^2 +(\epsilon_{n,k_z}-b)|\mu_{k_z}|^2\nonumber\\
 & + \mu^{\ast}_{k_z} \nu_{k_{z}}\Delta^{\ast} 
  + \mu_{k_z} \nu^{\ast}_{k_z} \Delta    \biggr)
  \gamma^{\dagger}_{n,-k_z,\downarrow} \gamma_{n,-k_z,\downarrow}  \Biggr],  \nonumber\\
  H_1&=\frac{eBL^2 L_z}{2\pi \hbar}  \sum_{n=0}^{n_{max}} \int \frac{d k_z}{2\pi}
 \Biggl[ \biggl(  2\epsilon_{n,k_z} \mu_{k_z}\nu_{k_z}+ \Delta^{\ast}\nu_{k_z}^2\nonumber\\
 &-\Delta \mu_{k_z}^{2}   \biggr)
  \gamma^{\dagger}_{n,k_z,\uparrow}\gamma^{\dagger}_{n,-k_z,\downarrow}  
  -\biggl(  2\epsilon_{n,k_z} \mu^{\ast}_{k_z}\nu^{\ast}_{k_z}+ \Delta \nu_{k_z}^{\ast 2}\nonumber\\
 & -\Delta^{\ast} \mu^{\ast 2}_{k_z}   \biggr)
   \gamma_{n,k_z,\uparrow} \gamma_{n,-k_z,\downarrow}   \Biggr], \nonumber\\
   E_g &=\frac{eBL^2 L_z}{2\pi \hbar}  \sum_{n=0}^{n_{max}} \int \frac{d k_z}{2\pi}
  \biggl(   2\epsilon_{n,k_z} |\nu_{k_z}|^2  -\Delta^{\ast} \mu^{\ast}_{k_z} \nu_{k_z} \nonumber\\
   &-\Delta \mu_{k_z} \nu^{\ast}_{k_z}   \biggr)
  +\frac{|\Delta|^2}{V_{int}}.
\end{align}
Since $H_1$ just includes off-diagonal terms in the $\gamma$ and $\gamma^{\dag}$ operators, we require it  to  vanish, leading to
\begin{align}
2\epsilon_{n,k_z} \mu_{k_z}\nu_{k_z}+ \Delta^{\ast}\nu_{k_z}^2-\Delta \mu_{k_z}^{2} =0,\nonumber\\
2\epsilon_{n,k_z} \mu^{*}_{k_z}\nu^{*}_{k_z}+ \Delta\nu_{k_z}^{*2 }-\Delta^{*} \mu_{k_z}^{*2} =0.
\end{align}

From  equation (3) and the two lines of equation (7), we obtain:
\begin{align}
  &|\mu_{k_z}|^2=\frac{1}{2}(1+\frac{\epsilon_{n,k_z}}{\xi_{n,k_z}}),\nonumber\\
  &|\nu_{k_z}|^2=\frac{1}{2}(1-\frac{\epsilon_{n,k_z}}{\xi_{n,k_z}}),
\end{align}
where $\xi_{n,k_z}=\sqrt{|\Delta|^2 + \epsilon^{2}_{n,k_z} }$.
Now the Hamiltonian, equation (1),  is diagonalized
\begin{align}
 H&=H_0+E_g \nonumber\\
  &=\frac{eBL^2 L_z}{2\pi \hbar}  \sum_{n=0}^{n_{max}} \int \frac{d k_z}{2\pi}
 \Biggl[ ( \xi_{n,k_z}+b )  \gamma^{\dagger}_{n,k_z,\uparrow} \gamma_{n,k_z,\uparrow} \nonumber\\
        &+( \xi_{n,k_z}-b ) \gamma^{\dagger}_{n,-k_z,\downarrow} \gamma_{n,-k_z,\downarrow}  \Biggr] \nonumber\\
        &+\frac{eBL^2 L_z}{2\pi \hbar}  \sum_{n=0}^{n_{max}} \int \frac{d k_z}{2\pi} (\epsilon_{n,k_z}-\xi_{n,k_z})+\frac{\Delta^2}{V_{int}},
\end{align}
and the gap $\Delta(T)$ in  equation (6) is given by
\begin{align}\label{Gap}
  \Delta(T)=V_{int}\sum_{n=0}^{n_{max}} \frac{eBL^2 L_z}{2\pi\hbar} \frac{d k_z}{2\pi} < a_{n,-k_{z},\downarrow} a_{n,k_{z},\uparrow} >.
\end{align}
Using the newly defined fermionic operators in equation (4), we may rewrite the above self-consistent  equation (10) as
\begin{align}
1=V_{int}\frac{eBL^2 L_z}{2\pi\hbar} \sum_{n=0}^{n_{max}} \int \frac{d k_z}{2\pi  }  
         \frac{  1- \frac{1}{1+ e^{\frac{ \xi_{n,k_{z}} +b  }{k_{B} T}}} -\frac{1}{1+ e^{\frac{ \xi_{n,k_{z}} -b  }{k_{B} T}}} }{2\xi_{n,k_z}}.  
\end{align}

When $\Delta(T)\rightarrow 0$, then $\xi_{n,k_z}\rightarrow |\epsilon_{n,k_z}|$, the elementary excitation of the superconducting state reduces to that of the normal state. We 
obtain the critical temperature $T_c(B)$ in equation (11) by setting $\Delta(T)=0$, yielding
\begin{align}
1=&V_{int}\frac{eBL^2 L_z}{4\pi^2 \hbar^2} \sqrt{\frac{m_z}{2}} \sum_{n} \int d E_{z}  
\frac{1}
{2\sqrt{E_z} \bigl( E_z+ ( n+\frac{1}{2})\hbar\omega -\mu_F  \bigr)} \nonumber\\
&\times\frac{ \sinh  \frac{  E_z+ ( n+\frac{1}{2})\hbar\omega -\mu_F }{k_B T_c}   }
{\cosh \frac{b}{k_B T_c}  + \cosh\frac{  E_z+ ( n+\frac{1}{2})\hbar\omega -\mu_F}{k_B T_c}  }.
\end{align}
         
\subsection{Electrons paired on  neighboring Landau levels}
The Hamiltonian of electrons paired on the neighboring LLs is
\begin{align}
H^{\prime}=&\sum_{n=0}^{n_{max}}\frac{eBL^2 L_z}{2\pi\hbar} \int \frac{d k_z}{2\pi}  \biggl[   (\epsilon_{n,k_z,}+b)  a^{\dagger}_{n,k_z,\uparrow}   
a_{n,k_z,\uparrow}\nonumber\\
                                                                         &+ (\epsilon_{n+1,k_z}-b) a^{\dagger}_{n+1,-k_z,\downarrow} a_{n+1,-k_z,\downarrow}   \biggr] 
                                                                         \nonumber\\
 &-\sum_{n=0}^{n_{max}}\frac{eBL^2 L_z}{2\pi\hbar} \int \frac{d k_z}{2\pi} \biggl(   \Delta^{\ast} a_{n,k_z,\uparrow} a_{n+1,-k_z,\downarrow} \nonumber\\
                                                                       & + \Delta   a^{\dagger}_{n+1,-k_z,\downarrow}a^{\dagger}_{n,k_z,\uparrow}           \biggr)
 +\frac{|\Delta|^2}{V_{int}},
\end{align}
in which $\Delta$ is
\begin{align}
   \Delta(T)=V_{int}\sum_{n-0}^{n_{max}} \frac{eBL^2 L_z}{2\pi\hbar} \frac{d k_z}{2\pi} < a_{n,k_{z},\uparrow} a_{n+1,-k_{z},\downarrow} >
\end{align}

We then employ a modified Bogoliubov transformation by defining the new fermionic operators:
\begin{align}
  \gamma_{n,k_z,\uparrow}&=\overline{\mu}_{k_z}a_{n,k_z,\uparrow}+\overline{\nu}_{k_z}a^{\dagger}_{n+1,-k_z,\downarrow},\nonumber\\
  \gamma_{n+1,-k_z,\downarrow}&=\overline{\mu}_{k_z}a_{n+1,-k_z,\downarrow}-\overline{\nu}_{k_z}a^{\dagger}_{n,k_z,\uparrow}; \nonumber\\
  \gamma^{\dagger}_{n,k_z,\uparrow}&=\overline{\mu}_{k_z}a^{\dagger}_{n,k_z,\uparrow}+\overline{\nu}_{k_z}a_{n+1,-k_z,\downarrow},\nonumber\\
  \gamma^{\dagger}_{n+1,-k_z,\downarrow}&=\overline{\mu}_{k_z}a^{\dagger}_{n+1,-k_z,\downarrow}-\overline{\nu}_{k_z}a_{n,k_z,\uparrow},
\end{align}
coefficients $\overline{\mu}_{k_z}$, $\overline{\nu}_{k_z}$ are real, and satisfy
\begin{align}
  \overline{\mu}_{k_z}^2+\overline{\nu}_{k_z}^2=1.
\end{align}
Then it is easy to obtain
\begin{align}
  a_{n,k_{z},\uparrow} &= \overline{\mu}_{k_z} \gamma_{n,k_z,\uparrow} - \overline{\nu}_{k_z}  \gamma^{\dagger}_{n+1,-k_{z},\downarrow},\nonumber\\
  a_{n+1,-k_{z},\downarrow}&=\overline{\mu}_{k_z}\gamma_{n+1,-k_z,\downarrow} + \overline{\nu}_{k_z} \gamma^{\dagger}_{n,k_z,\uparrow}; \nonumber\\
  a^{\dagger}_{n,k_{z},\uparrow}&=\overline{\mu}_{k_z} \gamma^{\dagger}_{n,k_z,\uparrow} - \overline{\nu}_{k_z}\gamma_{n+1,-k_{z},\downarrow},\nonumber\\
  a^{\dagger}_{n+1,-k_{z},\downarrow}&=\overline{\mu}_{k_z}\gamma^{\dagger}_{n+1,-k_z,\downarrow}+\overline{\nu}_{k_z}\gamma_{n,k_z,\uparrow}.
\end{align}
After substituting equation (17) into equation (13), the effective Hamiltonian becomes
\begin{align}
H^{\prime}=H_{0}^{\prime}+H_{1}^{\prime}+E^{\prime}_g ,
\end{align}
where
\begin{align}
  H_{0}^{\prime}=& \frac{eBL^2 L_z}{2\pi \hbar}  \sum_{n=0}^{n_{max}} \int \frac{d k_z}{2\pi}
 \Biggl[ \biggl(  (\epsilon_{n,k_z}+b )\overline{\mu}_{k_z}^2  \nonumber\\
 &-(\epsilon_{n+1,k_z}-b)\overline{\nu}_{k_z}^2+ 2\overline{\mu}_{k_z} \overline{\nu}_{k_{z}}\Delta     \biggr)
  \gamma^{\dagger}_{n,k_z,\uparrow} \gamma_{n,k_z,\uparrow}  \nonumber\\
  & +\biggl(  -(\epsilon_{n,k_z}+b )\overline{\nu}_{k_z}^2 +(\epsilon_{n+1,k_z}-b)\overline{\mu}_{k_z}^2\nonumber\\
  &+ 2\overline{\mu}_{k_z} \overline{\nu}_{k_{z}}\Delta     \biggr)
  \gamma^{\dagger}_{n+1,-k_z,\downarrow} \gamma_{n+1,-k_z,\downarrow}  \Biggr],  \nonumber\\
  H_{1}^{\prime}=&\frac{eBL^2 L_z}{2\pi \hbar}  \sum_{n=0}^{n_{max}} \int \frac{d k_z}{2\pi}
 \Biggl( (\epsilon_{n,k_z}+\epsilon_{n+1,k_z}) \overline{\mu}_{k_z}\overline{\nu}_{k_z} \nonumber\\
 &+ \Delta(\overline{\nu}_{k_z}^2- \overline{\mu}_{k_z}^{2})    \Biggr)
  (\gamma^{\dagger}_{n+1,-k_z,\downarrow}\gamma^{\dagger}_{n,k_z,\uparrow} \nonumber\\
  &+   \gamma_{n,k_z,\uparrow} \gamma_{n+1,-k_z,\downarrow}  ), \nonumber\\
   E_{g}^{\prime}=&\frac{eBL^2 L_z}{2\pi \hbar}  \sum_{n=0}^{n_{max}} \int \frac{d k_z}{2\pi}
   \biggl(   (\epsilon_{n,k_z} + \epsilon_{n+1,k_z} )\overline{\nu}_{k_z}^2 \nonumber\\ 
   &-2\Delta \overline{\mu}_{k_z} \overline{\nu}_{k_z}   \biggr)
  +\frac{|\Delta|^2}{V_{int}}.
\end{align}
Since $H_{1}^{\prime}$ includes off-diagonal terms in the $\gamma$ and $\gamma^{\dag}$ operators,  we require that it vanishes, implying that
\begin{align}
 (\epsilon_{n,k_z}+\epsilon_{n+1,k_z}) \overline{\mu}_{k_z}\overline{\nu}_{k_z} + \Delta(\overline{\nu}_{k_z}^2- \overline{\mu}_{k_z}^{2})  =0.
\end{align}
Combining  equations (16) and (20), we obtain:
\begin{align}
  &\overline{\mu}_{k_z}^2=\frac{1}{2}\Biggl(1+\frac{\epsilon_{n,n+1,k_z} }{\xi_{n,n+1k_z}  } \Biggr),\nonumber\\
  &\overline{\nu}_{k_z}^2=\frac{1}{2}\Biggl(1-\frac{\epsilon_{n,n+1,k_z} }{\xi_{n,n+1,k_z}  } \Biggr),
\end{align}
where $\xi_{n,n+1,k_z}=  \sqrt{ \Delta^2 +\epsilon_{n,n+1,k_z}^2}$, and $\epsilon_{n,n+1,k_z}= (\epsilon_{n,k_z} + \epsilon_{n+1,k_z})/2$.  Note that due to equation (20), 
$\overline{\mu}_{k_z}$ and $\overline{\nu}_{k_z}$ are functions of $\epsilon_{n,n+1,k_z}$ and $\Delta$.
Finally, the Hamiltonian for electrons paired on neighboring LLs is diagonalized and becomes 
\begin{align}
 H^{\prime}&=H_{0}^{\prime}+E_{g}^{\prime}\nonumber \\
  &=\frac{eBL^2 L_z}{2\pi \hbar}  \sum_{n=0}^{n_{max}} \int \frac{d k_z}{2\pi}
 \Biggl[ \biggl(  \xi_{n,n+1,k_z} -\epsilon_{n,n+1,k_z} + b   \biggr)  \nonumber\\
 & \times\gamma^{\dagger}_{n,k_z,\uparrow} \gamma_{n,k_z,\uparrow}\nonumber\\
   &+\biggl(  \xi_{n,n+1,k_z}  +\epsilon_{n,n+1,k_z} - b    \biggr)
  \gamma^{\dagger}_{n+1,-k_z,\downarrow} \gamma_{n+1,-k_z,\downarrow}  \Biggr]  \nonumber \\
  & +\frac{eBL^2 L_z}{2\pi \hbar}  \sum_{n=0}^{n_{max}} \int \frac{d k_z}{2\pi} (\epsilon_{n,n+1,k_z} -\xi_{n,n+1,k_z} )+\frac{\Delta^2}{V_{int}}.
\end{align}

Using the newly defined fermionic operators in equation (17), we may rewrite the self-consistent equation (14) as
\begin{align}
1&=V_{int}\frac{eBL^2 L_z}{2\pi\hbar} \sum_{n=0}^{n_{max}} \int \frac{d k_z}{2\pi  } \frac{1}{ 2\xi_{n,n+1,k_z}}\nonumber\\
&\times\biggl(
           1- \frac{1}{1+ e^{\frac{ \xi_{n,n+1,k_z} +b  }{k_{B} T}}} -\frac{1}{1+ e^{\frac{ \xi_{n,n+1,k_z} -b  }{k_{B} T}}}\biggr).  
\end{align}

When $\Delta(T)\rightarrow 0$, then $\xi_{n,n+1,k_z}\rightarrow |\epsilon_{n,n+1,k_z}|$, the elementary excitations of the superconducting state reduce to those of the normal 
state. We obtain the critical temperature $T_c(B)$ from  equation (23) by setting $\Delta(T)=0$, 
\begin{align}
1=&V_{int}\frac{eBL^2 L_z}{4\pi^2 \hbar^2} \sqrt{\frac{m_z}{2}} \sum_{n} \int d E_{z}  
\frac{1}
{2 \sqrt{E_z}\bigl( E_z+ ( n+1)\hbar\omega -\mu_F  \bigr)} \nonumber\\
&\times\frac{ \sinh  \frac{  E_z+ ( n+1)\hbar\omega -\mu_F }{k_B T_c}   }
{\cosh \frac{b}{k_B T_c}  + \cosh\frac{  E_z+ ( n+1)\hbar\omega -\mu_F}{k_B T_c}  }.
\end{align}

\section{Numerical results and Discussion}
\subsection{Numerical results}

\begin{figure}
\centering
{\includegraphics[width=0.45\textwidth]{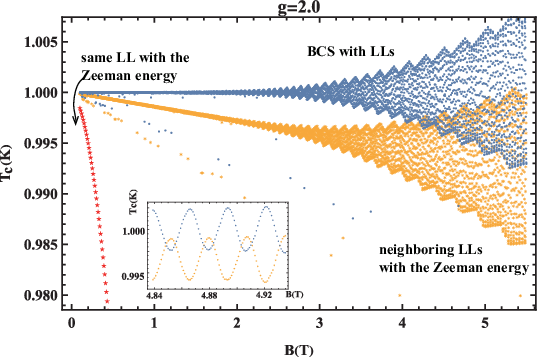}
\caption{The critical temperature $T_c(B)$ in magnetic fields for a fully isotropic system with $g=2$. The filled-blue circles represent $T_c(B)$ for the electrons paired on the same LL without the Zeeman energy, the yellow hexagrams represent electrons paired on neighboring LLs including the Zeeman energy, and the red five-point stars represent electrons paired on the same LLs including the Zeeman energy. The inset shows the oscillations of  $T_c(B)$  with an increasing magnetic field for electrons paired on the same LL without the Zeeman energy and on neighboring LLs with the Zeeman energy. Here we keep the value of the  interaction $V_{int}$ to be the same in the three cases.}
\label{fig2}}
\end{figure}

This work was done within the framework of the BCS theory, and we present  examples of $T_c(B)$ with the Fermi energy $\mu_F=0.1eV$ and the Debye frequency 
$\hbar\omega_D=0.01eV$, for which $T_c=1K$ in the BCS theory. We use the value of the attractive interaction $-V_{int}$ in the BCS theory, which satisfies $k_B 
T_c=\frac{2e^{\gamma}}{\pi}\hbar\omega_D \exp(-\frac{1}{N(0)V_{int}})$, $\gamma\approx0.5772$ is Euler's constant, $\frac{2e^{\gamma}}{\pi}\approx1.13$, and maintain $-V_{int}$ unchanged in different cases to eliminate the impact of the interaction upon $T_c$.  $N(0)$ is the electronic density of states at $\mu_{F}$. Here, we present our numerical results for the critical temperature in magnetic fields, initially assuming a fully isotropic system, which leads to $b=\frac{\hbar\omega}{2}$ and $g=2$. 

Figure 2 exhibits the critical temperature $T_{c}(B)$ in magnetic fields for a fully isotropic system with $g=2$. $T_c(B)$  is respectively represented  by the filled blue 
circles for the electrons paired on the same LL without the Zeeman energy, by the yellow hexagrams for electrons paired on neighboring LLs including the Zeeman energy,  and by 
the red five-pointed stars for electrons paired on the same LL including the Zeeman energy.  The critical temperature ($T_c$) for electrons paired on the same LL without the 
Zeeman energy oscillates around 1K in the standard BCS theory. For neighboring LLs with the Zeeman energy, it decreases in an oscillatory fashion as the magnetic field increases, as 
shown in the inset at high magnetic fields. The  depiction of $T_c(B)$ in figure 2 for electrons paired on the same LL with the Zeeman energy is incomplete, but our complete 
results are presented in figure 3.

\begin{figure}
\centering
{\includegraphics[width=0.45\textwidth]{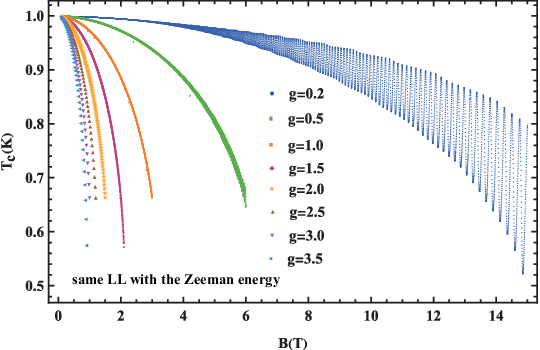}
\caption{The critical temperature $T_c(B)$ for electrons paired on the same LL in magnetic fields with different $g$ factors. The filled blue circles are for $g=0.2$, the filled green rectangles are for $g=0.5$, the filled yellow squares are for $g=1.0$, the filled raspberry diamonds are for $g=1.5$, the yellow five-pointed stars are for $g=2.0$, the brown filled triangles are for $g=2.5$, the magenta filled inverted triangles are for $g=3.0$, and the  blue filled left triangles are for $g=3.5$.}
\label{fig3}}
\end{figure}

As is commonly known, the response of an electron spin to a magnetic field can be affected by many issues, such as the effective masses of an electron in different 
directions \cite{Zhao-2021}, and the structure of the material \cite{Ebling-2014,Falson-2020}, etc. We studied the influence of the $g$ factor and the Zeeman energy on the critical temperature in a magnetic field, showing $T_c(B)$ for electrons paired on the same LLs in figure 3 with different $g$ factors. Notably, the Zeeman energy induces a decrease in $T_c$ with increasing magnetic field strength, and $T_c(B)$ drops sharply with an increasing $g$ factor.   This is due to the difference between electrons paired on the same LL with and without the Zeeman energy (see the filled blue circles in figure 2). $T_c(B)$ decreases in an oscillatory fashion when the $g$ factor is small.

\begin{figure}
\centering
{\includegraphics[width=0.45\textwidth]{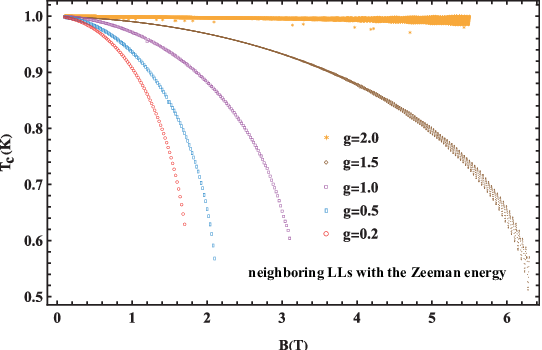}
\caption{The critical temperature $T_c(B)$ for electrons paired on  neighboring LLs in magnetic fields with different $g$ factors. The empty blue circles are for $g=0.2$, the empty blue rectangles are for $g=0.5$, the  empty magenta squares are for $g=1.0$, the  empty brown diamonds are for $g=1.5$, and the yellow hexagrams are for $g=2.0$.}
\label{fig3}}
\end{figure}

\begin{figure}
\centering
{\includegraphics[width=0.45\textwidth]{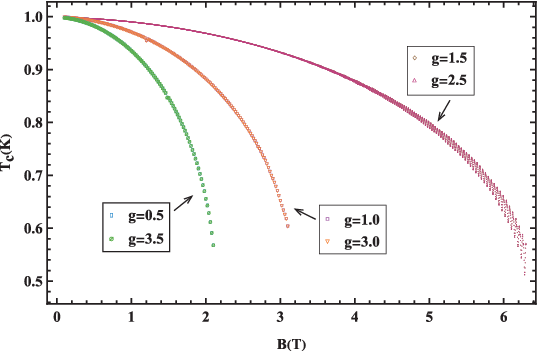}
\caption{The critical temperature $T_c(B)$ for electrons paired on neighboring LLs in magnetic fields with the Zeeman energy. The empty blue rectangles represent $T_c(B)$ 
for $g=0.5$, green $\emptyset$s for $g=3.5$, the empty magenta squares  are for $g=1.0$, the orange empty inverted triangles are for $g=3.0$, the empty brown diamonds are for $g=1.5$, and the raspberry empty triangles are for $g=2.5$.}
\label{fig5}}
\end{figure}

We further explored the effect of the $g$ factor (or the Zeeman energy) on the critical temperature for electrons paired on  neighboring LLs in varying magnetic fields, with the results shown in figures 4 and 5 for $g$ factors smaller than 4. As illustrated in figure 1, we restricted the $g$ factor to less than 4 to ensure that the number difference between adjacent LLs with opposite spins remains at 1. Figure 4 shows that $g=2$ gives the highest critical temperature in magnetic fields, which is also close to the $T_c$ in the BCS theory. The $g$ factor smaller or greater than 2 causes  $T_c$ to decrease in magnetic fields, and also $T_c(B)$ oscillates when $g$ is close to $2$. It is intriguing to observe that when the $g$-factor is in the range of $2\pm\delta$, where $\delta$ can encompass any number less than $2$, it results in comparable $T_c(B)$ values for electrons paired on neighboring LLs, as shown in figure 5.

\begin{figure}
\centering
{\includegraphics[width=0.45\textwidth]{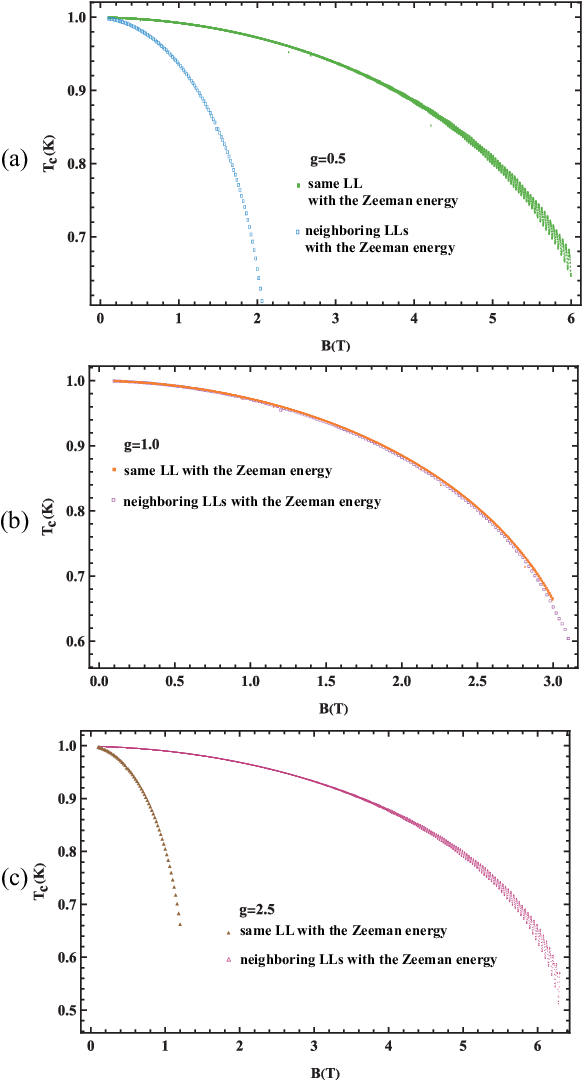}
\caption{The critical temperature $T_c(B)$ for electrons paired on the same and neighboring LLs in magnetic fields with the same $g$ value. The unfilled symbols denote results 
for electrons paired on the neighboring LLs, while the filled symbols indicate those paired on the same LLs. In Figs. 6(a), 6(b), and 6(c), we respectively display the results for $g$ values of 0.5 , 1.0, and 2.5.
}
\label{fig6}}
\end{figure}

To probe deeper into the effect of the Zeeman energy upon the critical temperature $T_c$ as a function of the magnetic field, we calculated and compared the $T_c(B)$ for electrons paired on the same and on neighboring LLs with the same $g$ value, presenting the results in figure 6. We present three cases with $g$ values of 0.5, 1.0, and 2.5, which provide a clear illustration that for $g$ less than 1, electrons paired on the same LL have the highest $T_c(B)$, whereas for $g$ greater than 1, those paired on  neighboring LLs have the highest $T_c(B)$. For the case of $g=1$, the $T_c(B)$ values for those two cases are nearly identical.

\subsection{Discussion}

In the present paper, we employed a fully quantum mechanical approach to examine the details of the effects of a high magnetic field upon the critical temperature of a Type-II 
superconductor, which are different from those obtained by treating the effects of the magnetic field semiclassically using the phase factor $\exp({\frac{ie}{\hbar}\int_{\bm{r}_1}^{\bm{r}_2}\bm{A}\cdot d\bm{r}})$. We expressed the energy of the paired electrons in an external magnetic field as the quantized energy 
levels $(n+\frac{1}{2})\hbar\omega+\frac{\hbar^{2}k_{z}^{2}}{2m_z}+\frac{g}{2}\mu_{B} \bm{\sigma}\cdot\bm{B}-\mu_F$, which have a high level of degeneracy for electrons paired in orbits perpendicular to the magnetic field. The quantized LLs, when influenced by the Zeeman energy $\frac{g}{2}\mu_{B} \bm{\sigma}\cdot\bm{B}$, allow the possibility of the electrons  forming Cooper pairs in the same or in neighboring LLs with opposite spins and momenta in the direction of the magnetic field. We employed the Bogoliubov-Valatin transformation to diagonalize the Hamiltonian (equations (1) and (13)) and to determine the critical temperature in magnetic fields by assuming that the gap vanishes in the self-consistent equations (10) and (14) at the critical temperature $T_c$. To eliminate the influence of the interaction upon the critical temperature, we consistently applied the same interaction values $-V_{int}$ in all of our analyses.

Our research demonstrates that in the absence of the Zeeman energy, the quantized LLs with electrons paired on the same LL with opposite spins result in oscillations of the 
critical temperature in strong magnetic fields around the $T_c$ predicted by the BCS theory, similar to the de-Haas-van Alphen effect. Conversely, in weak magnetic fields, $T_c$ remains nearly the same as for that described by the BCS theory, as shown  in figure 2. This occurs because for weak magnetic fields, the energy difference between 
neighboring LLs is quite small, making the situation nearly equivalent to that of free electrons forming Cooper pairs, as in the BCS theory. In a fully isotropic system, the Zeeman energy, denoted by $\pm\mu_{B}\bm{B}$ for the opposite spins of the electrons, leads to a splitting of the energy levels for electrons on the same LL, but electrons on 
neighboring LLs with opposite spins maintain the same energy levels, as illustrated in figure 1(c). Consequently, electrons paired on neighboring LLs exhibit a higher critical 
temperature $T_c$ in an oscillatory fashion compared to those paired on the same LL, and the critical temperature $T_c(B)$ for electrons paired on the same and on neighboring 
LLs both decrease as the magnetic field increases. It appears that electrons on neighboring LLs with opposite spins (or having the same energy) have a greater tendency to form 
Cooper pairs, which leads to a higher critical temperature than those paired on the same LL with opposite spins (having an energy difference) and demonstrates the quantized 
energy by exhbiting an oscillating $T_c(B)$ in the presence of magnetic fields. In fact, with increasing magnetic field strength, the number of LLs within the energy range 
$\mu_{F}\pm\hbar\omega_{D}$ decreases. Consequently, this reduction results in a decrease in the number of Cooper pairs, which subsequently leads to a decrease in the critical 
temperature with increasing magnetic field strength.

The $g$ factor is not always exactly 2, as it can be modified by many mechanisms. Numerous studies have sought to modify the $g$ factor to match experimental data 
\cite{Moehle,Zhang-2014}. Following this approach, we similarly investigate the Zeeman effect on the critical temperature of superconductors by varying the $g$ factor. We found that a smaller $ g$ factor for electrons paired on the same LL yields a higher critical temperature than does a larger $ g$ factor, and causes $T_c$ to decrease in an oscillatory fashion as the magnetic field strength increases, as illustrated in figure 3. In contrast, for electrons paired on  neighboring LLs, $g=2$ provides the highest critical temperature. Moreover, $T_c(B)$ displays an approximate symmetry around the $g$-value of 2, meaning the critical temperature is nearly identical for $g = 2 \pm\delta$, as demonstrated in figures 4 and 5. This symmetry around the $g$-factor of $2$ arises from the analogous distribution of LLs, for example, $g=1$ and $g=3$. The $(n-1, \uparrow)$ LL in $g=1$ has the same energy as the $(n, \downarrow)$ LL for $g=3$, and vice versa. Consequently, Cooper pairs formed on neighboring LLs have identical energies in these two cases, leading to identical critical temperatures in magnetic fields. For the $g$-factors between 1 and 4, pairing on neighboring LLs yields the highest $T_c(B)$. However, when $g$ is less than 1, pairing on the same LL results in the highest $T_c(B)$. This is attributed to the relatively small energy difference between electrons paired on the same LL with opposite spins when $g<1$. The same reasoning applies to $g$ values between 1 and 4. At $g=1$, the energy difference is identical for both pairing configurations.

Based on our numerical results, in the absence of the Zeeman energy, the critical temperature $T_c (B)$ for electrons paired on the same LL with opposite spins is consistently near to the BCS theory value. This might suggest that the BCS superconducting state can persist at any low temperature. However, this is not the case for electrons paired on the same or neighboring LLs when the Zeeman energy is included, as shown in the figures where $T_c(B)$ drops to a certain positive temperature. This indicates that the superconducting state can still exist at low temperatures, which is attributed to our assumption that the energy gap is zero at the critical temperature, and the excitation of superconductivity transforms into a normal state excitation. These results are different from the  semiclassical results of Gruenberg and Gunther. They claimed that there is a stable superconducting state at low temperature in the clean limit without the Zeeman energy, independent of the strength of the magnetic field. As for the oscillations of $T_c(B)$  displayed in our results, some early papers  using the semiclassical approximation also obtained similar conclusions \cite{GG-1966,Rajagopal-1966}.

Our present results were obtained in the clean limit. Impurities can 
significantly influence the properties of superconductors, such as the upper critical field. Similarly, impurities broaden the LLs, which in turn affects the critical 
temperature $T_c(B)$ in our paper. In future work, we plan to explore the effects of impurities upon the mixed pairing of electrons for the same and neighboring LLs on the 
critical temperature, the upper critical field, and the gap of Type-II superconductors.

\section{Conclusions}

In the present work, we investigated the impact of the discreteness of the LLs and the Zeeman energy on the critical field of superconductors in magnetic fields using a fully 
quantum mechanical approach based upon the BCS theory, rather than upon the semiclassical approach. Our findings revealed that Cooper pairs formed between electrons on the same 
LL with opposite spins lead to an oscillatory $T_c(B)$ around the BCS theory value when neglecting the spin response in high magnetic fields. For electrons paired on either the 
same or neighboring LLs with opposite spins, $T_c(B)$ decreases as the magnetic field increases due to the influence of the Zeeman energy. The behavior of $T_c(B)$ in 
magnetic fields suggests that electrons with close energy levels are more likely to form Cooper pairs, resulting in higher critical temperatures. Moreover, incorporating the 
Zeeman energy allows for the persistence of the superconducting state at low temperatures and high magnetic fields in the clean limit.

\begin{acknowledgments}
 This work was supported by the National Natural Science Foundation of China through Grant No. 11874083. 
\end{acknowledgments}

\end{document}